\documentclass[twocolumn,showpacs,amsmath,amssymb]{revtex4}
\input{epsf}
\usepackage{graphicx}
\usepackage{array}
\usepackage{longtable}
\usepackage{rotating}
\usepackage{booktabs,threeparttable}
\usepackage{color}

\usepackage{dcolumn,longtable}

\begin{document}

\title{\bf Critical nuclear charge and shape resonances for
the two-electron systems}

\author{Zong-Chao Yan}
\affiliation{ Department of Physics, University of New Brunswick, Fredericton, New Brunswick E3B 5A3 Canada}
\affiliation{State Key Laboratory of Magnetic Resonance and Atomic and Molecular Physics,
Wuhan Institute of Physics and Mathematics, Chinese Academy of Sciences, Wuhan 430071, China}
\affiliation{Center for Cold Atom Physics, Chinese Academy of Sciences, Wuhan 430071, China}
\author{Yew Kam Ho}
\affiliation{Institute of Atomic and Molecular Sciences, Academia Sinica, Taipei 10617, Taiwan, Republic of China}

\date{\today}

\begin{abstract}
The hydrogen negative ion H$^-$ is the simplest two-electron system that exists in nature. This system is not only important in
astrophysics but it also serves as an ideal ground to study electron-electron correlations. The peculiar balance of the correlations
between the two electrons with the interaction of electron-nucleus in H$^-$ makes this system to have only two bound states, one being the ground state $1s^2\,^{1}\!S^e$ and the other the doubly-excited metastable state $2p^2\,^{3}\!P^e$ embedded below the hydrogen $n=2$ threshold.
Here we report a calculation for the $2p^2\,^{3}\!P^e$ state of H$^-$ that yields the energy eigenvalue
$E=-0.125\,355\,451\,242\,864\,058\,376\,012\,313\,25(2)$, in atomic units.
Our result substantially improves the best available result by 16 orders of magnitude.
We further study the critical nuclear charge $Z_{\rm cr}$, the minimum value of nuclear charge $Z$ that is required to bind a nucleus and two electrons. Our determination of $Z_{\rm cr}$ for the $2p^2\,^{3}\!P^e$ state of two-electron systems
is $Z_{\rm cr}=0.994\,781\,292\,240\,366\,246\,3(1)$, corresponding to $1/Z_{\rm cr}= 1.005\,246\,085\,546\,985\,509\,4(1)$, which
improves the best published value of $Z_{\rm cr}$ by about 10 orders of magnitude.
We further investigate in a definitive way the unexplored regime of $Z < Z_{\rm cr}$ using the method of complex scaling
and establish precise shape resonance poles for the state of $2p^2\,^{3}\!P^e$ in the complex energy plane.

\end{abstract}


\maketitle

Two-electron atomic systems that exist in nature include the sequence H$^-$, He, Li$^+$,$\ldots$, which can be described
by the following single Hamiltonian in nonrelativistic and infinite nuclear mass limit (in atomic units throughout)
\begin{eqnarray}
\hat{H} &=& -\frac{1}{2}\nabla_1^2-\frac{1}{2}\nabla_2^2-\frac{Z}{r_1}-\frac{Z}{r_2}+\frac{1}{r_{12}}\,,
\label{eq1}
\end{eqnarray}
where ${\bf r}_1$ and ${\bf r}_2$ are the position vectors of electrons 1 and 2 and $r_{12}=|{\bf r}_1-{\bf r}_2|$.
There are two special bound states that always exist in the spectra of $\hat{H}$ for the above-mentioned sequence: one is
the ground state $1s^2\,^{1}\!S^e$, and the other is $2p^2\,^{3}\!P^e$ that is embedded between the $n=1$ and $n=2$ one-electron
ionization thresholds. These two states are truly bounded in the sense that they are stable against autoionization. Moreover,
for the first member in the two-electron sequence H$^-$, the state of $1s^2\,^{1}\!S^e$ is the only bound state that exists
below the $n=1$ ionization threshold, which has been proven mathematically~\cite{hill}; and the state of $2p^2\,^{3}\!P^e$ is the only bound state
embedded in the continuum that has been firmly established numerically
so far~\cite{Aashamar70,drake_prl_70,bhatia_pra_70,bunge_jcp_79,Banyard_jpb_92,bylicki_pra_03,kar_pra_jpb_2009}.
However, as $Z$ further reduces from $Z=1$, these two states
will eventually become unstable. For a given state, the minimum value of $Z$ that the Hamiltonian can still keep this state to be bounded
is called the critical nuclear charge $Z_{\rm cr}$ of the state of interest. A precise determination of the critical nuclear charge is
essential in understanding the analyticity of the function $E(Z)$ or $E(\lambda)$ in the complex $Z$ or $\lambda$ plane, where
$\lambda\equiv 1/Z$. In fact, for the ground state, $\lambda_{\rm s}\equiv1/Z_{\rm cr}$
is the radius of convergence for $E(\lambda)$ about $\lambda=0$ in the complex $\lambda$ plane, where $E(\lambda)$ is analytic
inside the circle $|\lambda|<\lambda_{\rm s}$ but has an essential singularity at
$\lambda=\lambda_{\rm s}\simeq 1.097\,660\,833$~\cite{baker_90,ivanov_pra_96,dubau_ivanov_jpb_98,drake_2014}.

For H$^-$, the energy eigenvalue of $1s^2\,^{1}\!S^e$ has been calculated to high precision~\cite{drake_2002,frolov_2007}.
The corresponding critical nuclear charge has been investigated by several authors, with the most precise calculation
of Estienne {\it et al.}~\cite{drake_2014} that sets a firm benchmark of 16-digit accuracy.
Atomic properties near and at the critical nuclear charge of $1s^2\,^{1}\!S^e$ have recently been studied by Grabowski and Burke~\cite{Grabowski}.
When $Z<Z_{\rm cr}$, the situation becomes
more complicate and subtle, as the bound state now crosses over the nearest ionization threshold and turns into a shape resonance.
Since a shape resonance lies in a scattering continuum, its energy, usually complex, cannot be defined directly from the original
Hamiltonian~(\ref{eq1}), but can be determined from its proper analytical continuation~\cite{dubau_ivanov_jpb_98}.
These resonances have been explored by Dubau and Ivanov~\cite{dubau_ivanov_jpb_98}, and Sergeev and Kais~\cite{sergeev_kais_ijqc_01} using the method of complex scaling. Comparing to the ground state, the analytic properties and resonant structure of the $2p^2\,^{3}\!P^e$ state
is much poorly understood. There are only two calculations available in the literature, to the best of our knowledge: one by
Br\"{a}ndas and Goscinski~\cite{brandas_ijqc} who give $1/Z_{\rm cr}=1.0048$ and one by Sergeev and Kais~\cite{sergeev_kais_jpa_99} who give
$1/Z_{\rm cr}=1.00524608$. Also to our knowledge, no studies have been reported to date concerning resonances in the region $Z<Z_{\rm cr}$
for the pure Coulomb case, notwithstanding that an investigation of shape resonances for the screened Coulomb case has been reported in Ref.~\cite{jiao_ho}.

The purpose of the present Letter is threefold. First, we present a substantially improved calculation on the
energy eigenvalue of the $2p^2\,^{3}\!P^e$ state in H$^-$.
Second, we establish a much more definitive determination for
the critical nuclear charge of this state. Finally we explore, for the first time, the unexplored region where $Z<Z_{\rm cr}$ and report
the $2p^2\,^{3}\!P^e$ resonant energies and widths using the method of complex scaling. In all these calculations, we employ Hylleraas
bases to describe the correlations between two electrons.

We use the following basis set in Hylleraas coordinates
\begin{eqnarray}
r_1^a r_2^b r_{12}^c \exp(-\alpha r_1-\beta r_2)\,Y_{\ell_1\ell_2}^{LM}(\hat{\bf r}_1,\hat{\bf r}_2)\,,
\label{eq2}
\end{eqnarray}
where $a \ge \ell_1$ and $b \ge \ell_2$,
\begin{eqnarray}
\small{
Y_{\ell_1\ell_2}^{LM}(\hat{\bf r}_1,\hat{\bf r}_2)=
\sum_{m_1m_2}
\langle \ell_1m_1\ell_2m_2|LM\rangle Y_{\ell_1}^{m_1}(\hat{\bf r}_1)Y_{\ell_2}^{m_2}(\hat{\bf r}_2)    }
\label{eq3}
\end{eqnarray}
is the common eigenfunction of the total angular momentum squared $\hat{L}^2$,  the $z$ component $\hat{L}_z$, and the parity operator with the
corresponding eigenvalues $L(L+1)$, $M$, and $(-1)^{\ell_1+\ell_2}$, respectively. The size of basis set is determined by including
all terms with $a+b+c\le \Omega$ with $\Omega$ being an integer. For the $2p^2\,^{3}\!P^e$ state of H$^-$,
we choose $\ell_1=\ell_2=L=1$. We include five blocks in the basis set each with different $(\alpha,\beta)$. In order to maintain maximum numerical stability of the basis set, we only include terms with $a\le b$. Starting from the second block,
all terms with $a=b$ are also excluded. Table~\ref{tab:I} shows a convergence study of the $2p^2\,^{3}\!P^e$
energy eigenvalue as the size of basis set increases. In the table, $R(\Omega)$ is the ratio of two successive differences that
can be used as a measure for the rate of convergence. We thus stop the calculation where $R(\Omega)\sim 1$.
The achieved accuracy of the extrapolated value of the energy is about
2 parts in $10^{28}$. Comparison with available published results indicates that the present calculation has dramatically improved the previous best result of Kar and Ho~\cite{kar_pra_jpb_2009} by about 16 orders of magnitude.

We now turn to the problem of finding the critical nuclear charge for the $2p^2\,^{3}\!P^e$ state. The general condition for this state being bound below the hydrogenic $n=2$ threshold is that $E(Z)\le -Z^2/(2n^2)$, or
$\epsilon(Z)\equiv -E(Z)/Z^2\ge 1/(2n^2)=0.125$. The basis set used in the variational calculation of the critical nuclear charge is
similar to that of the ground state except in the case of having five blocks in the basis, we only use two blocks.
The nonlinear parameters for the first block is approximately $(\alpha,\beta)\sim (0.5,0.01)$, and for the second block $(\alpha,\beta)\sim (0.7,0.5)$. Table~\ref{tab:II} lists the scaled energy $\epsilon(Z)$ for a series of calculations of different values of
$Z$. We use the arbitrary precision software package MPFUN, developed by Bailey~\cite{bailey}. The precision we set is 80 decimal
places. Our results in Table~\ref{tab:II} clearly shows that the critical nuclear charge $Z_{\rm cr}$ must be between
$0.994\,781\,292\,240\,366\,246\,0$ and $0.994\,781\,292\,240\,366\,246\,5$. We thus give an extrapolated value of $Z_{\rm cr}$ to be $0.994\,781\,292\,240\,366\,246\,3(1)$ that contains at least 18 significant digits after the decimal point. Our result is not only in agreement with the value of Sergeev and Kais~\cite{sergeev_kais_jpa_99} but also more precise than theirs by 10 more significant figures.

In order to explore the region of $Z<Z_{\rm cr}$, we apply the method of complex scaling~\cite{ho_complex}, by which the inter-particle coordinates
are transformed according to $r\rightarrow r \exp(i\theta)$ with the phase angle $\theta$ being real and positive. The Hamiltonian
(\ref{eq1}) is then transformed as $\hat{H}\rightarrow\hat{H}(\theta)=T \exp(-2i \theta)+V \exp(-i \theta)$, where $T$ and $V$ are
the kinetic and potential operators respectively. A diagonalization of the transformed Hamiltonian yields complex energy
eigenvalues of the form $E_{\rm res}=E_r-i\Gamma/2$, where $E_r$ is the resonance position and $\Gamma$ the resonance width.
In search of a resonance, a resonance pole is the one that exhibits the most stabilized behavior of the energy with respect to
changes of $\theta$ and nonlinear parameters in the basis set. Table~\ref{tab:III} shows numerical results for some selected
complex resonance energies with different $Z<Z_{\rm cr}$. It should be mentioned that we have used two different types of basis sets:
the Hylleraas and the configuration interaction (CI) and the results are consistent up to $4 \sim 5$ digits for the shape resonances
in Table~\ref{tab:III}.

\begin{figure}
\includegraphics[width=17cm, height=15cm]{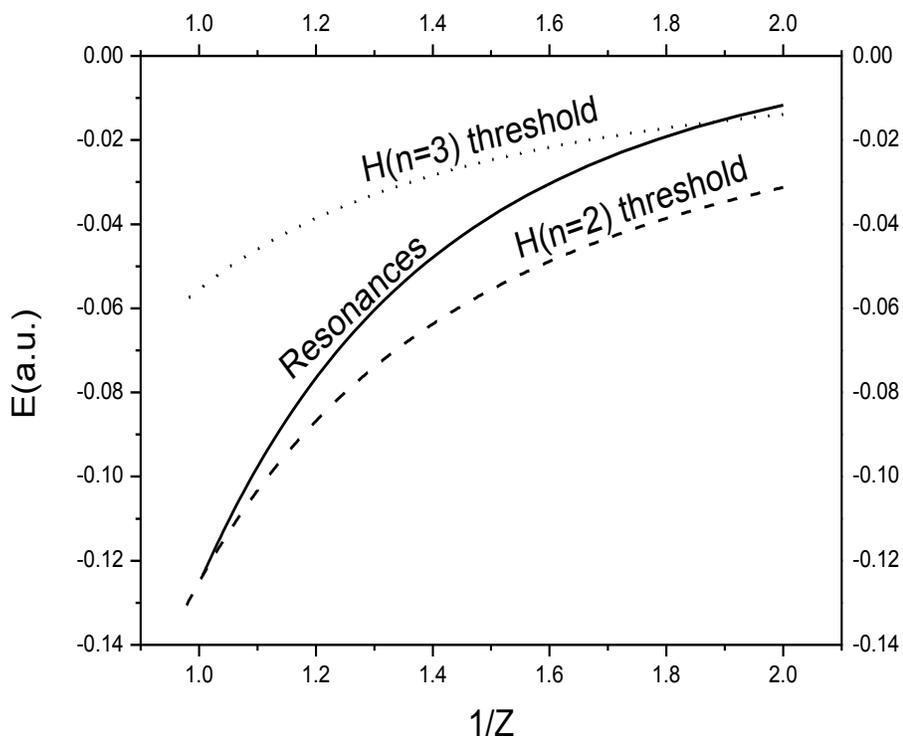}
\caption{(Color online) Resonance energy vs $1/Z$ for the $2p^2\,^{3}\!P^e$ state in two-electron systems.}
\label{fig1}
\end{figure}

Figure~\ref{fig1} is the plot of resonance energy versus $1/Z$ for this shape resonance. It is seen that the shape resonance starts to appear just above the two-body H $(n=2)$ threshold as $Z$ is first decreased ($1/Z$ is increased) below $Z_{\rm cr}$. As $Z$ is decreased further, the shape resonance moves away from the two-body H $(n=2)$ threshold. When $1/Z$ is increased further to near the region around 1.8, the shape resonance starts to approach the H $(n=3)$ threshold and it even crosses over the H $(n=3)$ threshold when $1/Z$ is increased to around 1.9. We stop at $1/Z \sim 2.0$, because as the resonance is approaching the total ionization limit $E=0$, it is getting more difficult to obtain convergence in resonance calculations.

\begin{figure}
\includegraphics[width=17cm, height=23cm]{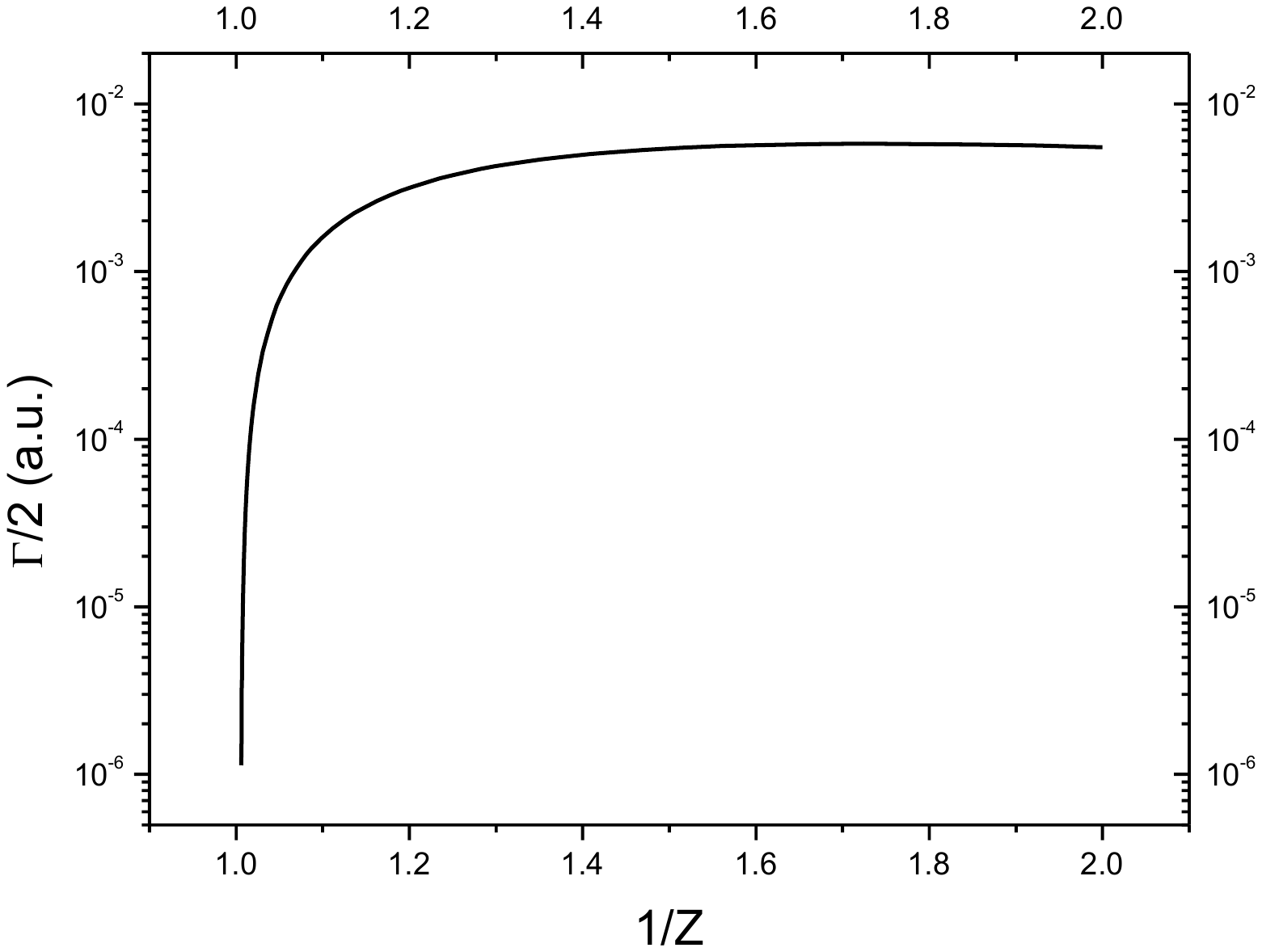}
\caption{(Color online) Half resonance width vs $1/Z$ for the $2p^2\,^{3}\!P^e$ state in two-electron systems, in logarithmic scale.}
\label{fig2}
\end{figure}

Figure~\ref{fig2} is the plot of the half-width $\Gamma/2$ versus $1/Z$ in logarithmic scale.
It is seen that the width increases very rapidly when the bound state first turns to a shape resonance as its energy crosses over the H $(n=2)$ threshold. The parent state of this shape resonance is hence the excited $2p$ state of the two-body atom, together with the other electron having a $p$-wave character (two spin-half $p$-wave electrons are coupled to form a $^3P^e$ state). Now when the resonance crosses the $n=2$ threshold from a bound state to a shape resonance at $Z_{\rm cr}$, the $p$-wave electron will experience a repulsive angular momentum barrier, through which the electron can tunnel out, resulting in a cross-over from a bound state to a shape resonance. Furthermore, as the shape resonance moves away from the threshold when $Z$ is decreased further, the thickness of the potential barrier, through which the electron tunnels out, becomes narrower, leading to a shorter lifetime for the autoionization process, hence broadening the width, a consequence of the uncertainty principle. The physics of such a phenomenon is similar to the electron tunneling in field ionization when an atom is placed under an external electric field.

One can see from Fig.~\ref{fig2} that $\Gamma/2$ reaches a maximum value near $1/Z \sim 1.7$, and then it starts to decrease somewhat as $1/Z$ further increases. This can be explained as follows. As shown in Fig.~\ref{fig1},
when $Z$ is decreased below $Z_{\rm cr}$, a bound state becomes a shape resonance lying above the H $(n=2)$ threshold. As $Z$ is decreased further, the shape resonance is moving away from the H $(n=2)$ threshold and approaching the H $(n=3)$ threshold. As such, the resonance is now gaining some Feshbach-type character. In other words, at this range of $1/Z$, the resonance is a mixture of a shape resonance, with the H $(n=2)$ as the parent state, and a Feshbach resonance, with the H $(n=3)$ as the parent state. The autoionization process is a combination of tunneling effect, as described earlier, and Feshbach-type configuration-interaction effect as one electron decaying from the $n=3$ shell to the $n=2$ shell, with the other electron being autoionized away. Usually, for a Feshbach resonance the width would decrease as it is moving toward its parent ($n=3$, in this case) threshold. So for our special case, the total width is a competition between the ``increasing character" due to tunneling effect and the ``decreasing character" due to the Feshbach configuration-interaction effect. Apparently, as the resonance approaches the H $(n=3)$ threshold, the Feshbach resonance effect would become more pronounced, resulting in the overall decreasing trend for the total width around this $1/Z$ region.

In conclusion, we have established two new benchmark results for the $2p^2\,^{3}\!P^e$ state, {\it i.e}, the energy eigenvalue of H$^-$ and the critical nuclear charge for the two-electron isoelectronic sequence. We have also explored in a definitive way the resonance poles as the nuclear charge is below its critical value. Our results will be valuable in studying the analytical structure of $E(\lambda)$ in the complex $\lambda$ plane.

ZCY was supported by NSERC of Canada and by the CAS/SAFEA International Partnership Program for Creative Research Teams.
YKH was supported by Ministry of Science and Technology of Taiwan.
Research support from the computing facilities of SHARCnet and ACEnet is gratefully acknowledged.

\clearpage

\begin{table}
\caption{Convergence of the nonrelativistic energy for the $2p^2\,^{3}\!P^e$ state of H$^-$ with infinite nuclear mass.
In atomic units.}
\label{tab:I}
\begin{tabular}{lclc}
\hline\hline
\multicolumn{1}{l}{$\Omega$} & \multicolumn{1}{c}{No. of terms} & \multicolumn{1}{c}{$E(\Omega)$} &
\multicolumn{1}{c}{$R(\Omega)$} \\
\hline
14 & 1064  &  --0.125\,355\,451\,242\,864\,057\,032\,738\,135\,921\,31  &    \\
15 & 1316  &  --0.125\,355\,451\,242\,864\,058\,265\,245\,521\,423\,53  &    \\
16 & 1604  &  --0.125\,355\,451\,242\,864\,058\,368\,324\,462\,549\,58  &11.95    \\
17 & 1932  &  --0.125\,355\,451\,242\,864\,058\,375\,223\,510\,548\,96  &14.94    \\
18 & 2301  &  --0.125\,355\,451\,242\,864\,058\,375\,939\,082\,805\,06  &9.64    \\
19 & 2715  &  --0.125\,355\,451\,242\,864\,058\,376\,004\,981\,517\,50  &10.85    \\
20 & 3175  &  --0.125\,355\,451\,242\,864\,058\,376\,011\,372\,600\,17  &10.31    \\
21 & 3685  &  --0.125\,355\,451\,242\,864\,058\,376\,012\,166\,845\,94  &8.04    \\
22 & 4246  &  --0.125\,355\,451\,242\,864\,058\,376\,012\,296\,758\,16  &6.11    \\
23 & 4862  &  --0.125\,355\,451\,242\,864\,058\,376\,012\,311\,685\,94  &8.70    \\
24 & 5534  &  --0.125\,355\,451\,242\,864\,058\,376\,012\,312\,844\,98  &12.87    \\
25 & 6266  &  --0.125\,355\,451\,242\,864\,058\,376\,012\,313\,076\,82  &4.99    \\
26 & 7059  &  --0.125\,355\,451\,242\,864\,058\,376\,012\,313\,169\,71  &2.49    \\
27 & 7917  &  --0.125\,355\,451\,242\,864\,058\,376\,012\,313\,202\,85  &2.80    \\
28 & 8841  &  --0.125\,355\,451\,242\,864\,058\,376\,012\,313\,222\,51  &1.68    \\
29 & 9835  &  --0.125\,355\,451\,242\,864\,058\,376\,012\,313\,231\,80  &2.11    \\
30 & 10900 &  --0.125\,355\,451\,242\,864\,058\,376\,012\,313\,237\,21  &1.71 \\
Extrap.  & &  --0.125\,355\,451\,242\,864\,058\,376\,012\,313\,25(2)    &         \\
Aashamar~\cite{Aashamar70} (1970)                       & &--0.125\,325 & \\
Drake~\cite{drake_prl_70} (1970)                        & &--0.125\,350  &         \\
Bhatia~\cite{bhatia_pra_70} (1970)                      & &--0.125\,354\,705\,1 & \\
J\'auregui and Bunge~\cite{bunge_jcp_79} (1979)         & &--0.125\,354\,716\,6 & \\
                                                        & &--0.125\,355\,08(10) (extrap.)& \\
Banyard {\it et al.}~\cite{Banyard_jpb_92} (1992)       & &--0.125\,353\,6 &\\
Bylicki and Bednarz~\cite{bylicki_pra_03} (2003)        & &--0.125\,355\,451\,24 &\\
                                                        & &--0.125\,355\,453\,06 (extrap.) & \\
Kar and Ho~\cite{kar_pra_jpb_2009} (2009)               & &--0.125\,355\,451\,242 & \\
\hline\hline
\end{tabular}
\end{table}

\clearpage

\begin{table}
\caption{Scaled energy eigenvalue $\epsilon(Z)\equiv-E(Z)/Z^2$ for the state of $2p^2\,^{3}\!P^e$ as a function of $Z$. The size of basis set
in each calculation is 11900. In atomic units.}
\label{tab:II}
\begin{tabular}{lc}
\hline\hline
\multicolumn{1}{c}{$Z$} & \multicolumn{1}{c}{$\epsilon(Z)$} \\
\hline
 0.994\,781\,292\,240\,370\,000\,0& 0.125\,000\,000\,000\,000\,228\,829\,28        \\
 0.994\,781\,292\,240\,367\,000\,0& 0.125\,000\,000\,000\,000\,048\,404\,64        \\
 0.994\,781\,292\,240\,366\,400\,0& 0.125\,000\,000\,000\,000\,009\,865\,61        \\
 0.994\,781\,292\,240\,366\,250\,0& 0.125\,000\,000\,000\,000\,000\,230\,86        \\
 0.994\,781\,292\,240\,366\,247\,0& 0.125\,000\,000\,000\,000\,000\,038\,16        \\
 0.994\,781\,292\,240\,366\,246\,5& 0.125\,000\,000\,000\,000\,000\,006\,05        \\
 0.994\,781\,292\,240\,366\,246\,0& 0.124\,999\,999\,999\,999\,999\,973\,93        \\
Extrapolation: &\\
0.994\,781\,292\,240\,366\,246\,3(1)  &\\
0.994\,781\,29\footnotemark[1]& \\
\hline\hline
\end{tabular}
\footnotetext[1]{Ref.~\cite{sergeev_kais_jpa_99}.}
\end{table}


\begin{table}
\caption{Selected shape resonance poles for the state $2p^2\,^{3}\!P^e$ for various $Z$ values when $Z < Z_{\rm cr}$.
In the table, $a[b]\equiv a\times 10^b$. In atomic units.}
\label{tab:III}
\begin{tabular}{lcc}
\hline\hline
\multicolumn{1}{c}{$Z$} & \multicolumn{1}{c}{$E_r$} &
\multicolumn{1}{c}{$\Gamma/2$} \\
\hline
 0.993 &  $-1.231466[-1]$ &    $4.9685[-6]$ \\
 0.990 &  $-1.222194[-1]$ &    $2.8275[-5]$ \\
 0.985 &  $-1.206725[-1]$ &    $8.7928[-5]$ \\
 0.980 &  $-1.191255[-1]$ &    $1.6185[-4]$ \\
 0.970 &  $-1.160375[-1]$ &    $3.3449[-4]$ \\
 0.960 &  $-1.129645[-1]$ &    $5.2682[-4]$ \\
 0.940 &  $-1.068834[-1]$ &    $9.4119[-4]$ \\
 0.920 &  $-1.009096[-1]$ &    $1.3716[-3]$ \\
 0.900 &  $-9.505911[-2]$ &    $1.8031[-3]$ \\
 0.850 &  $-8.103785[-2]$ &    $2.8364[-3]$ \\
 0.800 &  $-6.795992[-2]$ &    $3.7552[-3]$ \\
 0.700 &  $-4.486793[-2]$ &    $5.1245[-3]$ \\
 0.600 &  $-2.607694[-2]$ &    $5.7485[-3]$ \\
 0.500 &  $-1.170528[-2]$ &    $5.5065[-3]$ \\
 \hline\hline
 \end{tabular}
\end{table}

\clearpage


\begin{thebibliography}{99}

\bibitem{hill} R. N. Hill, Phys. Rev. Lett. {\bf 38}, 643 (1977).

\bibitem{Aashamar70} K. Aashamar, Nucl. Instrum. Meth. {\bf 90}, 263 (1970).


\bibitem{drake_prl_70} G. W. F. Drake, Phys. Rev. Lett. {\bf 24}, 126 (1970).

\bibitem{bhatia_pra_70} A. K. Bhatia, Phys. Rev. A {\bf 2} 1667 (1970)

\bibitem{bunge_jcp_79} R. J\'auregui and C. F. Bunge, J. Chem. Phys. {\bf 71}, 4611 (1979).

\bibitem{Banyard_jpb_92} K. E. Banyard, D. R. T. Keeble, and G. W. F. Drake, J. Phys. B {\bf 25}, 3405 (1992).

\bibitem{bylicki_pra_03} M. Bylicki and E. Bednarz, Phys. Rev. A {\bf 67}, 022503 (2003).

\bibitem{kar_pra_jpb_2009} S. Kar and Y. K. Ho, J. Phys. B {\bf 42}, 185005 (2009).

\bibitem{baker_90} J. D. Baker, D. E. Freund, R. N. Hill, and J. D. Morgan III,
Phys. Rev. A {\bf 41}, 1247 (1990).

\bibitem{ivanov_pra_96} I. A. Ivanov, Phys. Rev. A {\bf 54}, 2792 (1996).

\bibitem{dubau_ivanov_jpb_98} J. Dubau and I. A. Ivanov, J. Phys. B {\bf 31}, 3335 (1998).

\bibitem{drake_2014} C. S. Estienne, M. Busuttil, A. Moini, and G. W. F. Drake, Phys. Rev. Lett. {\bf 112}, 173001 (2014).

\bibitem{drake_2002} G. W. F. Drake, M. M. Cassar, and R. A. Nistor, Phys. Rev. A {\bf 65}, 054501 (2002).

\bibitem{frolov_2007} A. M. Frolov, J. Phys. A  {\bf 40}, 6175 (2007).

\bibitem{Grabowski} P. E. Grabowski and K. Burke, Phys. Rev. A {\bf 91}, 032501 (2015).

\bibitem{sergeev_kais_ijqc_01} A. V. Sergeev and S. Kais, Int. J. Quantum Chem. {\bf 82}, 255 (2001).

\bibitem{brandas_ijqc} E. Br\"{a}ndas and O. Goscinski, Int. J. Quantum Chem. {\bf 6}, 59 (1972).

\bibitem{sergeev_kais_jpa_99} A. V. Sergeev and S. Kais, J. Phys. A {\bf 32}, 6891 (1999).

\bibitem{jiao_ho} L. G. Jiao and Y. K. Ho, J. Quant. Spectrosc. Radiat. Transfer {\bf 144}, 27 (2014).

\bibitem{bailey} http://crd-legacy.lbl.gov/$\sim$~dhbailey/mpdist/

\bibitem{ho_complex} Y. K. Ho, Phys. Rep. {\bf 99}, 1 (1983).

\end{thebibliography}
\end{document}